\RequirePackage{lineno}
\documentclass[twocolumn,showpacs,amsmath,amssymb,superscriptaddress,nofootinbib]{revtex4-1}
\usepackage{amsmath}
\usepackage{amssymb}
\usepackage{graphicx}
\usepackage{dcolumn}
\usepackage{bm}

\usepackage{amsfonts}
\usepackage{keyval,graphicx}
\usepackage{textcomp,wasysym}
\usepackage{multirow}

\usepackage[titletoc]{appendix}
\usepackage{lineno}
\usepackage{tabularx}
\usepackage{color}

\usepackage[colorlinks, linkcolor=blue, anchorcolor=blue, citecolor=blue]{hyperref}                        

\newcommand{\jpsi}{J/\psi}
\newcommand{\g}{\gamma}
\newcommand{\pip}{\pi^+}
\newcommand{\pim}{\pi^-}

\newcommand{\fz}{f_{0}(980)}
\newcommand{\etap}{\eta^{\prime}}
\newcommand{\kap}{K^+}
\newcommand{\kam}{K^-}
\newcommand{\ar}{\rightarrow}
\newcommand{\GeV}{GeV/$c^2$}
\newcommand{\MeV}{MeV/$c^2$}
\newcommand{\Y}{Y(2175)}
\newcommand{\BR}{\mathcal{B}}
\newcommand{\X}{X}
\newcommand{\I}{\uppercase\expandafter{\romannumeral1}}
\newcommand{\II}{\uppercase\expandafter{\romannumeral2}}

\begin{document}

\title{\boldmath Observation and study of the decay $J/\psi\rightarrow\phi\eta\eta'$}

\author{
\small
M.~Ablikim$^{1}$, M.~N.~Achasov$^{10,d}$, S. ~Ahmed$^{15}$, M.~Albrecht$^{4}$, M.~Alekseev$^{56A,56C}$, A.~Amoroso$^{56A,56C}$, F.~F.~An$^{1}$, Q.~An$^{53,43}$, J.~Z.~Bai$^{1}$, Y.~Bai$^{42}$
% ...
\\
\vspace{0.2cm}
(BESIII Collaboration)\\
\vspace{0.2cm} {\it
$^{1}$ Institute of High Energy Physics, Beijing 100049, People's Republic of China\\
% ...
$^{4}$ Bochum Ruhr-University, D-44780 Bochum, Germany\\
% ...
$^{10}$ G.I. Budker Institute of Nuclear Physics SB RAS (BINP), Novosibirsk 630090, Russia\\
% ...
$^{15}$ Helmholtz Institute Mainz, Johann-Joachim-Becher-Weg 45, D-55099 Mainz, Germany\\
% ...
$^{42}$ Southeast University, Nanjing 211100, People's Republic of China\\
$^{43}$ State Key Laboratory of Particle Detection and Electronics, Beijing 100049, Hefei 230026, People's Republic of China\\
% ...
$^{53}$ University of Science and Technology of China, Hefei 230026, People's Republic of China\\
% ...
$^{56}$ (A)University of Turin, I-10125, Turin, Italy; (B)University of Eastern Piedmont, I-15121, Alessandria, Italy; (C)INFN, I-10125, Turin, Italy\\
% ...
$^{d}$ Also at the Novosibirsk State University, Novosibirsk, 630090, Russia\\
% ...
}
}

\begin{abstract}


We report the observation and study of the decay $J/\psi\rightarrow\phi\eta\eta'$ using $1.3\times{10^9}$ $J/\psi$ events collected with the BESIII detector. Its branching fraction, including all possible intermediate states, is measured to be $(2.32\pm0.06\pm0.16)\times{10^{-4}}$. 
We also report evidence for a structure, denoted as $X$, in the $\phi\eta'$ mass spectrum in the $2.0-2.1$ GeV/$c^2$ region. 
Using two decay modes of the $\eta'$ meson ($\gamma\pi^+\pi^-$ and $\eta\pi^+\pi^-$), 
a simultaneous fit to the $\phi\eta'$ mass spectra is performed. 
Assuming the quantum numbers of the $X$ to be $J^P = 1^-$, 
its significance is found to be 4.4$\sigma$, with a mass and width of $(2002.1 \pm 27.5 \pm 21.4)$ MeV/$c^2$ and $(129 \pm 17 \pm 9)$ MeV, respectively, and a product branching fraction $\mathcal{B}(J/\psi\rightarrow\eta{}X)\times{}\mathcal{B}(X\rightarrow\phi\eta')=(9.8 \pm 1.2 \pm 1.7)\times10^{-5}$. 
Alternatively, assuming $J^P = 1^+$, the significance is  3.8$\sigma$, with a mass and width of $(2062.8 \pm 13.1 \pm 7.2)$ MeV/$c^2$ and $(177 \pm 36 \pm 35)$ MeV, respectively, and a product branching fraction $\mathcal{B}(J/\psi\rightarrow\eta{}X)\times{}\mathcal{B}(X\rightarrow\phi\eta')=(9.6 \pm 1.4 \pm 2.0)\times10^{-5}$. 
The angular distribution of $J/\psi\rightarrow\eta{}X$ is studied and the two $J^P$ assumptions of the $X$ cannot be clearly distinguished due to the limited statistics. 
In all measurements the first uncertainties are statistical and the second systematic.
\end{abstract}


\maketitle



\section{INTRODUCTION}

Exotic hadrons, e.g., glueballs, hybrid states and multiquark states, are allowed in the framework of Quantum Chromodynamics (QCD), but no conclusive evidence for them has yet been found in the light hadron sector. The decay $\jpsi\rightarrow{VPP}$ (where $V$ denotes vector and $P$ denotes pseudoscalar) is an ideal probe to study light hadron spectroscopy and to search for new hadrons. There have been theoretical~\cite{c_introduction_the_1, c_introduction_the_2, c_introduction_the_3, c_introduction_the_4} and experimental~\cite{c_introduction_exp_MARKIII, c_introduction_exp_DM2, c_introduction_exp_omega_pipi, c_introduction_exp_omega_kk, c_introduction_exp_phi_pipi_kk, c_introduction_exp_kstr_k_pi} studies performed, which have mainly been focused on the $V$ recoil system to search for exotic hadrons. The $P$ recoil system, on the other hand, could also be utilized to do a similar study. For example, the $\Y$, denoted as $\phi(2170)$ by the Particle Data Group (PDG)~\cite{c_introduction_pdg}, was confirmed in the process $J/\psi\rightarrow\eta{Y(2175)}, Y(2175)\ar\phi{f_0(980)}$ by BESII~\cite{c_introduction_bes} and BESIII~\cite{c_introduction_besiii}. 
Searching for its decay to the $\phi\etap$ state provides valuable input for understanding its nature~\cite{c_introduction_ssg}. 
The decay $\jpsi\ar\phi\eta\etap$ has not been studied before, and could aid in our understanding of $\jpsi$ decay mechanisms and offers an opportunity to study possible intermediate states.

In this article, we report the observation and study of the decay $\jpsi\ar\phi\eta\etap$ using $(1310.6 \pm 7.0)\times{10^6}$ $\jpsi$ events~\cite{c_data_samples} collected with the BESIII detector. Its branching fraction, including all possible intermediate states, is measured. We also report evidence for a structure denoted as $\X$ in the $\phi\etap$ mass spectrum in the $2.0-2.1$ \GeV\ region. The mass and width of this structure, as well as the product branching fraction $\BR(\jpsi\ar\eta\X)\times{}\BR(\X\ar\phi\etap)$, 
are measured. 
The $\phi$ meson is reconstructed through its $K^{+}K^{-}$ decay mode, $\eta$ through $\gamma\gamma$, and $\etap$ through both $\g\pip\pim$ and $\eta\pip\pim$ (with the $\eta\ar\g\g$), denoted as mode \uppercase\expandafter{\romannumeral1} and mode \uppercase\expandafter{\romannumeral2}, respectively.


\section{BESIII EXPERIMENT AND MONTE CARLO SIMULATION}

The BESIII detector is a magnetic spectrometer~\cite{c_besiii_detector} located at the Beijing Electron Position Collider (BEPCII)~\cite{c_bepcii}. The cylindrical core of the BESIII detector consists of a helium-based multilayer drift chamber (MDC), a plastic scintillator time-of-flight system (TOF), and a CsI(Tl) electromagnetic calorimeter (EMC), which are all enclosed in a superconducting solenoidal magnet providing a 1.0 T (0.9 T in 2012) magnetic field. The solenoid is supported by an octagonal flux-return yoke with resistive plate counter muon identifier modules interleaved with steel. The acceptance of charged particles and photons is 93$\%$ over 4$\pi$ solid angle. The charged-particle momentum resolution at 1 GeV/$c$ is 0.5$\%$, and the $dE/dx$ resolution is 6$\%$ for electrons from Bhabha scattering. The EMC measures photon energies with a resolution of 2.5$\%$ (5$\%$) at 1 GeV in the barrel (end cap) region. The time resolution of the TOF barrel part is 68~ps, while that of the end cap part is 110~ps. 

Simulated data samples produced with a {\sc geant4}-based~\cite{c_MC_GEANT4} Monte Carlo~(MC) package, including the geometric description of the BESIII detector and the detector response, are used to determine the detection efficiency and to estimate the backgrounds. The simulation of the $e^+e^-$ collisions includes the beam energy spread and initial state radiation~(ISR) and is modeled using the generator {\sc kkmc}~\cite{c_MC_KKMC}. 
The inclusive MC sample consists of the production of the $\jpsi$ resonance and the continuum processes incorporated in {\sc kkmc}~\cite{c_MC_KKMC}. The known decay modes are modeled with {\sc evtgen}~\cite{c_MC_BesEvtGen} using branching fractions taken from the PDG~\cite{c_introduction_pdg}, and the remaining unknown decays from the charmonium states with {\sc lundcharm}~\cite{c_MC_LUNDCHARM}. Final state radiation~(FSR) from charged final state particles is incorporated with the {\sc photos} package~\cite{c_MC_PHOTOS}.


\section{Event selection and data analysis}

Charged tracks are reconstructed from hits in the MDC. We select four charged tracks with net charge zero in the polar angle range $|\cos\theta|<0.93$, and require their points of closest approach to the $e^+e^-$ interaction point to be within $\pm10$ cm in the beam direction and $1$ cm in the plane perpendicular to the beam direction. The $dE/dx$ and TOF measurements are combined to form particle identification (PID) confidence levels for the $\pi$, $K$ and $p$ hypotheses. We require that one $K^+K^-$ pair and one $\pip\pim$ pair are identified. A vertex fit that assumes the $\pip\pim\kap\kam$ tracks all come from a common vertex is applied. 

Photons are reconstructed from electromagnetic showers in the EMC. At least three photons are required for mode \uppercase\expandafter{\romannumeral1} and four for mode \uppercase\expandafter{\romannumeral2}. The minimum energy for showers to be identified as photons in the barrel region ($|\cos\theta|<0.8$) is 25 MeV, and in the end caps  ($0.86<|\cos\theta|<0.92$) is 50 MeV. Showers out of the above regions are poorly reconstructed and not used in this analysis. To suppress showers from charged particles, a photon must be separated by at least 10 degrees from the nearest charged track. EMC cluster timing requirements suppress electronic noise and energy deposits unrelated to this event.

\begin{figure}[htbp]
\centering
{\includegraphics[width=0.4\textwidth]{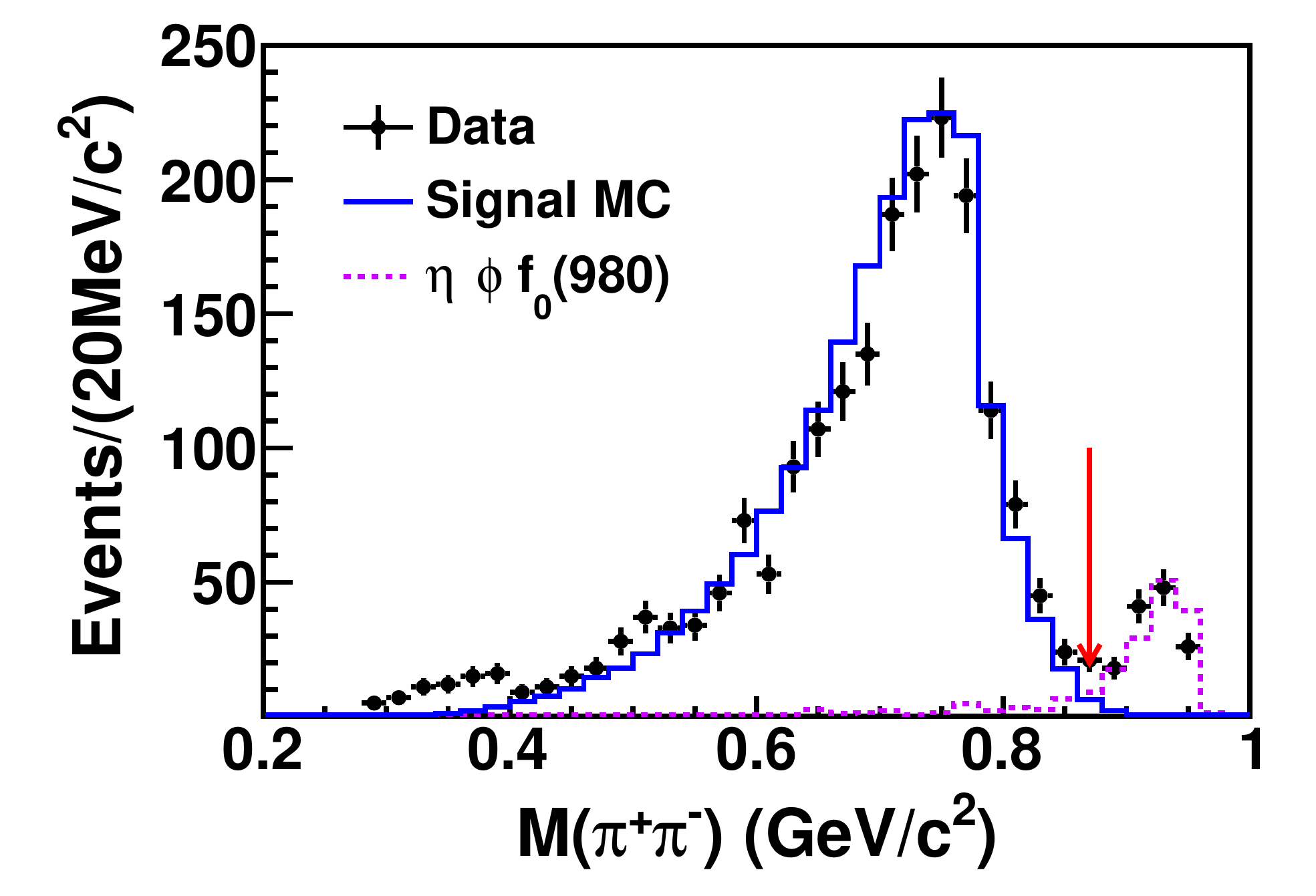}}
\caption{\label{m_2pi_f1} The $M(\pip\pim)$ distribution for mode \uppercase\expandafter{\romannumeral1}, where dots with error bars are experimental data, the (blue) solid histogram shows the signal MC simulation, the (violet) dotted histogram shows the background from the $\jpsi\ar\eta\phi f_0(980)$ process, and the arrow represents the mass requirement.}
\end{figure}

Four-constraint (4C) kinematic fits are applied to all combinations of photons, and only the combination with the smallest $\chi^2_{\rm 4C}$ is kept.  
We only keep those events with $\chi^2_{\rm 4C}\le{40}$ for mode \uppercase\expandafter{\romannumeral1} and $\chi^2_{\rm 4C}\le{80}$ for mode \uppercase\expandafter{\romannumeral2}. To suppress background events containing $\pi^0$'s, those events with the invariant mass of any photon pair within a $\pi^0$ mass window [$0.12\le{}M(\g\g)\le{}0.15$ \GeV] are rejected. For mode \uppercase\expandafter{\romannumeral1}, the combination with the smallest value of $\delta^2_1=[M(\gamma_1\gamma_2)-m_{\eta}]^2/\sigma_\eta^2+[M(\gamma_3 \pi^+ \pi^-)-m_{\eta'} ]^2/\sigma_{\eta'}^2$ is used to assign photons to the $\eta$ and $\eta'$. Here $m_{\eta}$ and $m_{\eta'}$ are the nominal $\eta$ and $\eta'$ masses~\cite{c_introduction_pdg}, respectively; $\sigma_{\eta}$ and $\sigma_{\eta'}$ are the mass resolutions determined from signal MC simulation. Mass windows for the $\eta$, $\phi$ and $\etap$ mesons are (in \GeV) $0.522\le{}M(\g\g)\le{}0.573$, $1.010\le{}M(K^+K^-)\le{}1.030$ and $0.936\le{}M(\g\pip\pim)\le{}0.979$. $M(\pip\pim)$ is required to be less than 0.87 \GeV\ to suppress the background from the $\jpsi\ar\eta\phi\fz$ process as shown in Fig.~\ref{m_2pi_f1}. For mode \uppercase\expandafter{\romannumeral2}, we use the combination with the smallest $\delta^2_2=[M(\gamma_1\gamma_2)-m_{\eta}]^2/\sigma_\eta^2+[M(\gamma_3\gamma_4)-m_{\eta}]^2/\sigma_{\eta}^2$ for the best $\eta$ meson combination; the $\eta$ for which $M(\pi^+\pi^-\eta)$ is closest to $m_{\eta'}$ is attributed to the candidate decaying from the $\eta'$. Mass windows for the $\eta$, $\phi$ and $\etap$ mesons are (in \GeV) $0.509\le{}M(\g\g)\le{}0.586$, $1.010\le{}M(K^+K^-)\le{}1.030$ and $0.920\le{}M(\eta\pip\pim)\le{}0.995$. 

\begin{figure*}[htbp]
\centering
{\includegraphics[width=0.8\textwidth]{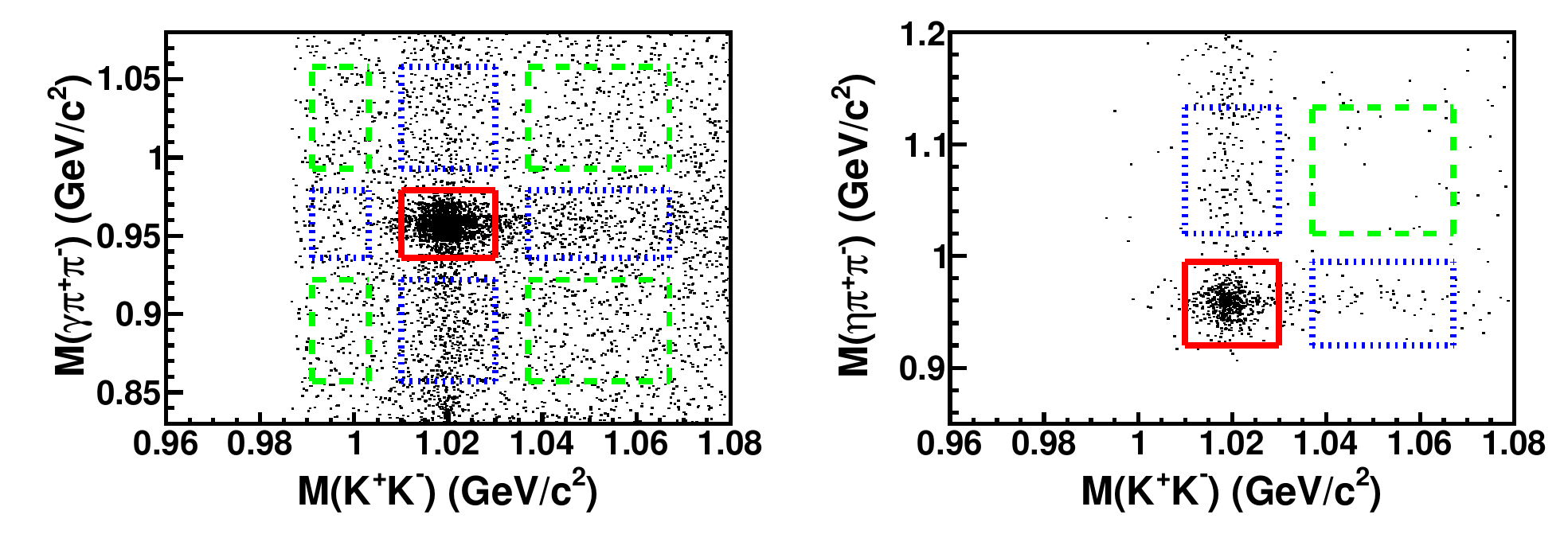}}
    \put(-357,115){\large (a)}
    \put(-154,115){\large (b)}
\caption{\label{2D_4_plots} Distributions of $M(\g\pip\pim)$ versus $M(K^+K^-)$ for mode \uppercase\expandafter{\romannumeral1} (a) and $M(\eta\pip\pim)$ versus $M(K^+K^-)$ for mode \uppercase\expandafter{\romannumeral2} (b), where the (red) solid rectangles show the signal regions; the (blue) dotted and (green) dashed rectangles represent the 2D sidebands.}
\end{figure*}

Figure \ref{2D_4_plots} shows the distributions of $M(\g\pip\pim)$ versus $M(K^+K^-)$ for mode \uppercase\expandafter{\romannumeral1} and $M(\eta\pip\pim)$ versus $M(K^+K^-)$ for mode \uppercase\expandafter{\romannumeral2}. The background inferred from the $\eta$ sidebands is negligible according to both the study of the data and the corresponding inclusive MC samples for $\jpsi$ decays. The non-$\phi$ and/or non-$\etap$ backgrounds are determined by the weighted sums of the horizontal and vertical sidebands with the entries in the diagonal sidebands subtracted to compensate for the double counting of background components. The different sidebands are illustrated in Fig.~\ref{2D_4_plots} and the weighting factors are obtained from the 2-dimensional (2D) fits to the mass spectra of $M(\g\pip\pim)$ versus $M(K^+K^-)$ and $M(\eta\pip\pim)$ versus $M(K^+K^-)$. The $\phi$ and $\etap$ meson signals are seen clearly in both modes. The three body decay $\jpsi\ar\phi\eta\etap$ is thus established, which is the first observation of this decay.

\section{Measurement of $\BR(\jpsi\ar\phi\eta\etap)$}

The branching fraction for $\jpsi\ar\phi\eta\etap$, including all possible intermediate states, is measured. Following the procedure in Ref.~\cite{c_3body_br_efficiency}, the regions of $M^2(\phi\etap)$ versus $M^2(\phi\eta)$ are divided into $40\times{40}$ areas (each area is tagged by $i$ and $j$) and the numbers of events ($n_{\rm data}^{ij}$), non-$\phi$ and/or non-$\etap$ background ($n_{\rm bkg}^{ij}$) and efficiency ($\epsilon_{ij}$) are obtained individually in each area. Then $\BR(\jpsi\ar\phi\eta\etap)$ is determined by 
\begin{eqnarray}
\begin{split}
\mathcal{B}=\frac{N_{\rm corr}}{N_{J/\psi}\mathcal{B}(\eta\rightarrow2\gamma)\mathcal{B}(\phi\rightarrow K^+K^-)\mathcal{B}_{\etap}},
\end{split}
\end{eqnarray}
where $N_{\rm corr}$ is the efficiency-corrected number of signal events and is determined from $N_{\rm corr}=\Sigma_{ij}[(n_{\rm data}^{ij}-n_{\rm bkg}^{ij})/\epsilon_{ij}$]; $N_{\jpsi}$ is the total number of $\jpsi$ events~\cite{c_data_samples}; $\BR$ is the PDG branching fraction~\cite{c_introduction_pdg}; $\BR_{\etap}$ is $\BR(\etap\ar\g\pip\pim)$ for mode \I\ and $\BR(\etap\ar\eta\pip\pim)\times{}\BR(\eta\ar\g\g)$ for mode \II. The total signal yield after background subtraction is $1684\pm48$ for mode \I{} and $510\pm25$ for mode \II{}; $\BR(\jpsi\ar\phi\eta\etap)$ is determined to be $(2.31\pm0.07)\times10^{-4}$ for mode \uppercase\expandafter{\romannumeral1} and $(2.34\pm0.12)\times10^{-4}$ for mode \uppercase\expandafter{\romannumeral2}. The uncertainties are statistical only. 
The weighted average~\cite{c_combine_3body} of the results for the two $\etap$ decay modes is $(2.32 \pm 0.06 \pm 0.16) \times 10^{-4}$, after taking into account the correlations between uncertainties from the two modes, as denoted with asterisks in Table~\ref{t_sys_2}.

The systematic uncertainties in $\BR(\jpsi\ar\phi\eta\etap)$ measurements are shown in Table \ref{t_sys_2}. The uncertainties from MDC tracking and PID efficiencies are established to be 1.0$\%$ per pion/kaon in Refs.~\cite{c_uncertainty_track, c_uncertainty_PID}. The uncertainty related to photon detection is determined to be 0.6$\%$ per photon in Ref.~\cite{c_uncertainty_photon}. The uncertainties associated with the 4C kinematic fit are studied with the track parameter correction method~\cite{c_uncertainty_kinematic} and the differences between the efficiencies with and without corrections are regarded as uncertainties; the influence of the $\chi^2_{\rm 4C}$ requirement is also considered in the uncertainty determination. The sideband regions of the $\phi$ and $\etap$ mesons are shifted by 1$\sigma$ (the nominal width of signal region corresponds to 3$\sigma$), and the effects on the results are assigned as uncertainties. The uncertainties from mass windows are determined by smearing the mass spectra from MC simulation to compensate for the differences between the resolutions from data and MC; the differences between efficiencies before and after smearing are taken as uncertainties. The influences of finite MC statistics are taken into account. The uncertainties due to quoted branching fractions and number of $J/\psi$ events are from the PDG~\cite{c_introduction_pdg} and Ref.~\cite{c_data_samples}, respectively. The uncertainties from the 2D binning method are obtained by changing the numbers of areas in the $\BR(\jpsi\ar\phi\eta\etap)$ determination. The total systematic uncertainties are obtained by summing all contributions in quadrature, assuming they are independent.  

    \begin{table} [tbh]
    \centering
    \caption{Systematic uncertainties in $\mathcal{B}(J/\psi\rightarrow\phi\eta\eta')$. The correlated sources between the two $\etap$ decay modes are denoted with asterisks.}
    \label{t_sys_2}
    \begin{tabular} {lcc}
    \hline
    \hline
    Sources                                               & Mode \I~($\%$) & Mode \II~($\%$) \\
    \hline
    MDC tracking*   $\quad$$\qquad$$\qquad$$\qquad$$\,$              & 4.0 & 4.0 \\
    PID*                                                      & 4.0 & 4.0 \\
    Photon detection*                                 & 1.8 & 2.4 \\
    Kinematic fit                                         & 2.5 & 1.1 \\
    Sideband regions                                & 0.1 & 0.3 \\    
    Mass window for $\eta$                      & 0.5 & 0.7 \\
    Mass window for $\phi$                       & 0.9 & 1.0 \\
    Mass window for $\eta'$                      & 0.7 & 0.6 \\
    MC statistics                                        & 0.6 & 0.9 \\
    Branching fractions*                              & 2.1 & 2.1 \\
    Number of $J/\psi$*                              & 0.6 & 0.6 \\ 
    2D binning                                           & 3.9 & 2.2 \\
    Total                                                     & 8.0 & 7.2 \\ 
    \hline
    \hline
    \end{tabular}
    \end{table}


\section{Study of an intermediate state in the $\phi\etap$ mass spectrum}

\begin{figure*}[htbp]
\centering
{\includegraphics[width=0.8\textwidth]{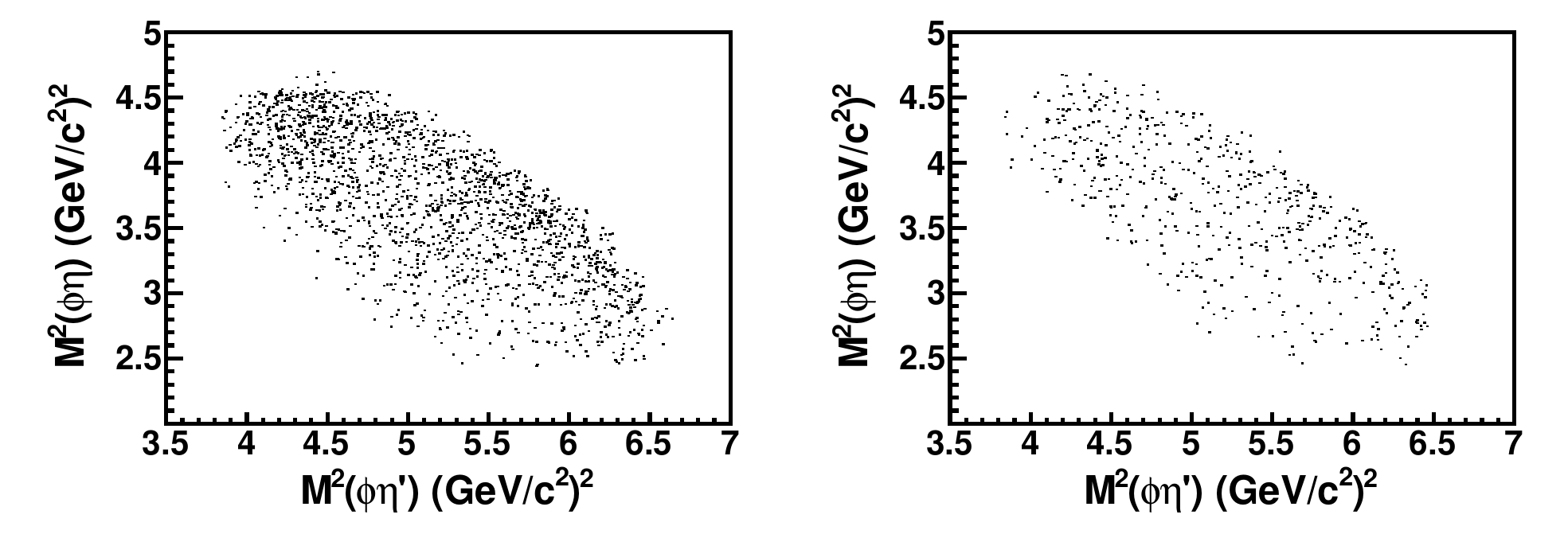}}
    \put(-238,115){\large (a)}
    \put(-34,115){\large (b)}
    \caption{\label{dalitz_2modes} Dalitz plots for modes \uppercase\expandafter{\romannumeral1} (a) and \uppercase\expandafter{\romannumeral2} (b).}
\end{figure*}

Figure \ref{dalitz_2modes} shows Dalitz plots for modes \uppercase\expandafter{\romannumeral1} and \uppercase\expandafter{\romannumeral2}. Both have concentrations of events with $M^{2}(\phi\etap)$ values near 4.5 $(\rm GeV/\it{c}^{\rm2})^{\rm2}$. There are also diagonal bands in both modes corresponding to the process $\jpsi\ar\phi{}f_0(1500), f_0(1500)\ar\eta\etap$ according to studies of the MC samples. Apart from these, no other structures are evident. 

\subsection{Simultaneous fit}

With the assumption that there is an $\X$ structure in the $\phi\etap$ mass spectrum
in the $2.0-2.1$ \GeV\ region, corresponding to the clusters near 4.5
$(\rm GeV/\it{c}^{\rm2})^{\rm2}$ visible in Fig.~\ref{dalitz_2modes},
a simultaneous fit is performed on the $\phi\etap$ mass spectra for
modes \I~and \II. Since the spin-parity
  value ($J^P$) of the structure could affect the relative orbital angular
  momenta between the decay products of $\jpsi\ar\eta\X$ and
  $\X\ar\phi\etap$, the fits with two different assumptions on the
  $J^P$ value are both performed. However, due to the limited
  statistics, they cannot clearly be distinguished. In the simultaneous
  fits, the interference between the structure and the direct decay $J/\psi\ar\phi\eta\etap$ is not considered.
 
Assuming the $J^P$ value of the structure to be $1^-$, the signal component is parameterized by
%
\begin{equation}
\label{f_sig}
\begin{split}
(|\frac{1}{m^2-M^2+iM\Gamma/c^2}|^2\times{(pq)^3\times\epsilon})\otimes{R},
\end{split}
\end{equation} 
where $m$ is the reconstructed mass of the $\phi\etap$ system; $M$ and
$\Gamma$ are the mass and width of the structure in the
constant-width relativistic Breit-Wigner (BW) function; the P-wave
phase space (PHSP) factor $(pq)^3$ is considered in the partial width,
where $p$ is the $\phi$ momentum in the $\phi\etap$ rest frame, and $q$ is
the $\eta$ momentum in the $\jpsi$ rest frame; $\epsilon$ denotes the
efficiency and $R$ is the double-Gaussian resolution function, both of
which are determined from a signal MC simulation. The mass and width of
the BW function are allowed to float but are constrained to be the
same for both modes; the signal ratio of the two modes is fixed based
on PDG $\etap$ branching fractions~\cite{c_introduction_pdg} and
MC-determined efficiencies. The total signal yield for the two modes is
allowed to float in the fit. The background components consist of
non-resonant $\phi\eta\etap$,
$\jpsi\ar\phi{}f_0(1500), f_0(1500)\ar\eta\etap$ and non-$\phi$ and/or
non-$\etap$ processes. For the non-resonant $\phi\eta\etap$ process,
the line shapes are derived from the MC
simulation of $\jpsi\ar\phi\eta\etap$ process generated according to PHSP, and the ratio of background numbers for the two modes is
fixed, similar to the signal case. For
$\jpsi\ar\phi{}f_0(1500), f_0(1500)\ar\eta\etap$ background, whose influence on the structure is small, the shapes are from MC simulation;
$\BR(\jpsi\ar\phi{}f_0(1500))\times{\BR(f_0(1500)\ar\pi\pi)}$ and
$\BR(\jpsi\ar\phi{}f_0(1500))\times{\BR(f_0(1500)\ar{}K\bar{K})}$ from
BESII~\cite{c_introduction_exp_phi_pipi_kk}, together with $\BR(f_0(1500)\ar\pi\pi)$,
$\BR(f_0(1500)\ar{}K\bar{K})$ and $\BR(f_0(1500)\ar\eta\etap)$ from
the PDG~\cite{c_introduction_pdg}, are used to obtain the expected
number of $f_0(1500)$, and the background number is
fixed to the expected value. The non-$\phi$ and/or non-$\etap$ backgrounds are
determined from the 2D sidebands of the $\phi$ and $\etap$ mesons as
shown in Fig.~\ref{2D_4_plots}.

\begin{figure*}[htbp]
\centering
{\includegraphics[width=0.8\textwidth]{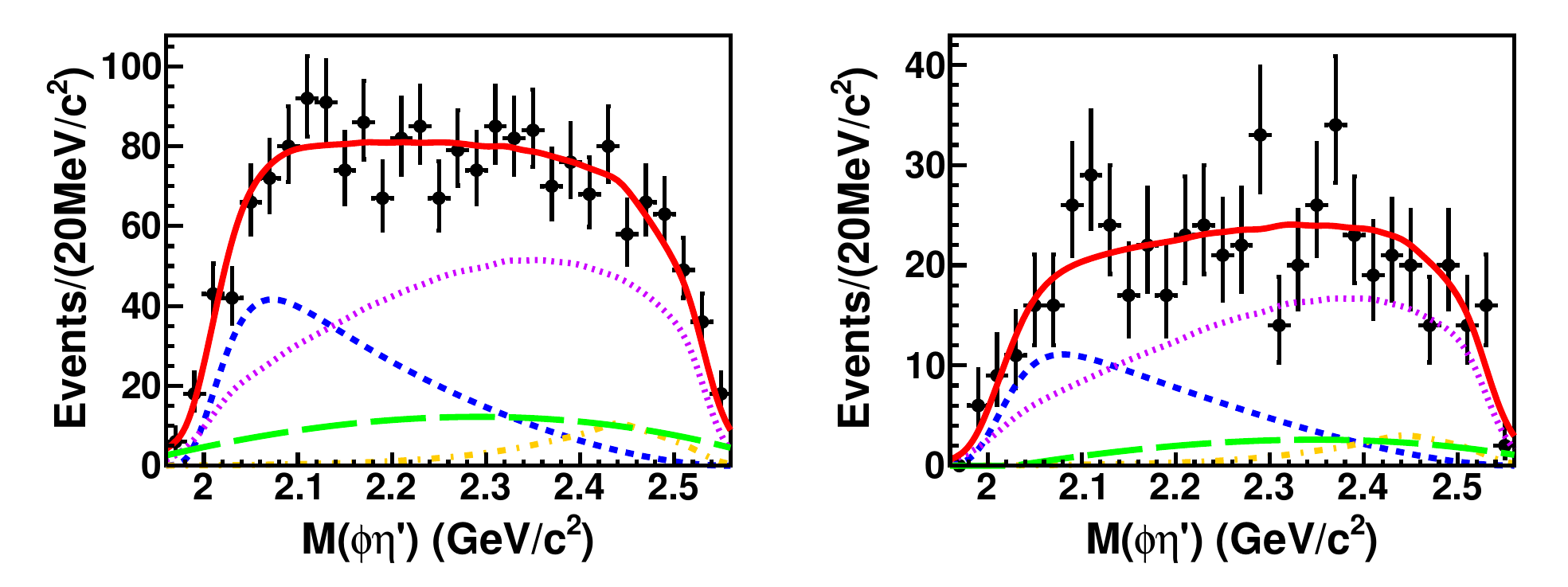}}
    \put(-238,125){\large (a)}
    \put(-34,125){\large (b)}
\caption{\label{fit_simultaneous} Results of the simultaneous fit with the $1^-$ assumption for modes  \uppercase\expandafter{\romannumeral1} (a) and \uppercase\expandafter{\romannumeral2} (b). Dots with error bars are experimental data and the (red) solid curves show the fit model. The (blue) dashed curves are the signal component. The (violet) dotted curves show the background from the $J/\psi\rightarrow\phi\eta\eta'$ PHSP process. The (orange) dot-dashed curves represent the background from the $J/\psi\rightarrow\phi f_0(1500), f_0(1500)\rightarrow\eta\eta'$ process. The (green) long-dashed curves show the non-$\phi$ and/or non-$\etap$ backgrounds.}
\end{figure*}

Figure~\ref{fit_simultaneous} shows the results of the simultaneous
fit, where the mass and width of the structure are determined to be
$(2002.1\pm27.5)$ \MeV\ and $(129\pm17)$ MeV,
respectively. The log-likelihood value is
  15591.8, with a goodness-of-fit value of $\chi^2/\rm{d.o.f.}$
  of $20.98/26=0.81$ for mode \I\ and $25.97/26=1.00$ for
  mode \II. The statistical significance of the new structure is
calculated to be larger than 10$\sigma$, determined from the change of
the log-likelihood values and the numbers of free parameters in the
fits with and without the inclusion of the structure. After
smearing the likelihood curve with the Gaussian-distributed systematic
uncertainties (Table~\ref{t_sys}), the significance is evaluated to be
4.4$\sigma$. Many checks have been done to make sure that none of the
possible background contributions could produce peaking backgrounds
in the $2.0-2.1$ \GeV\ region in the $\phi\etap$ mass spectrum. A comparison between
data and MC also indicates no significant structures in the
$\phi\eta$ mass spectrum.

\begin{figure*}[htbp]
\centering
{\includegraphics[width=0.8\textwidth]{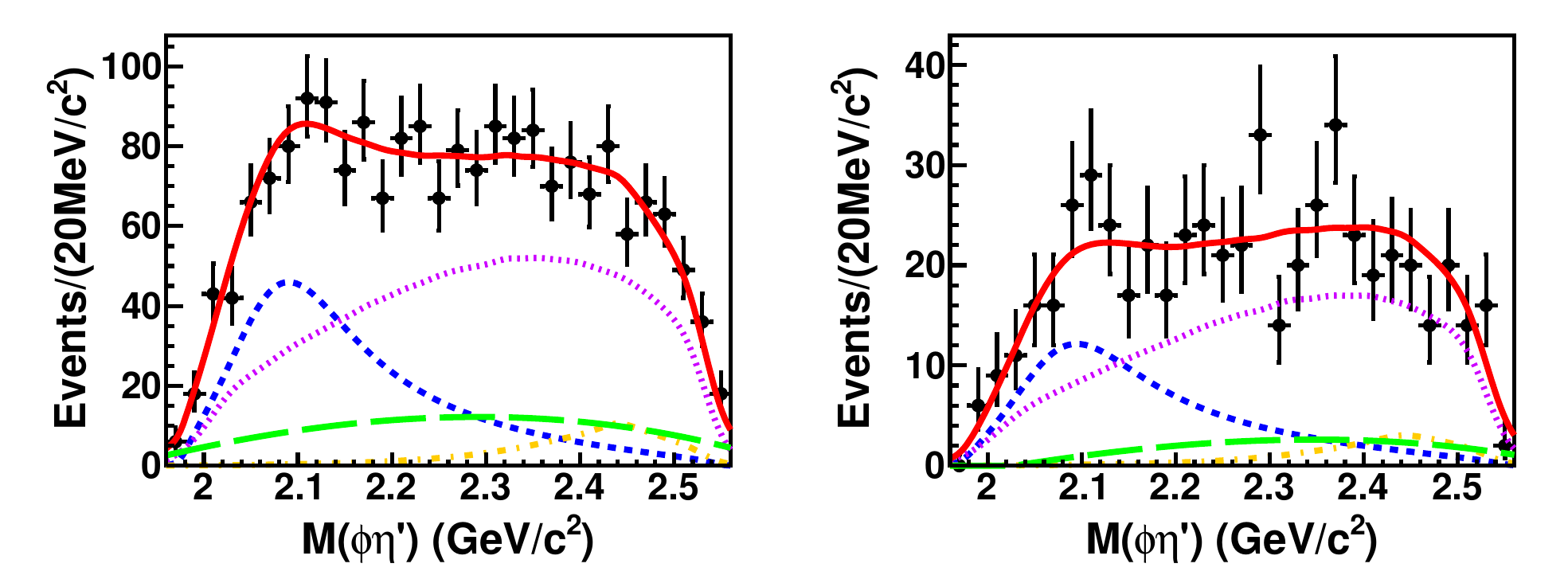}}
    \put(-238,125){\large (a)}
    \put(-34,125){\large (b)}
\caption{\label{fit_simultaneous_ss} Results of the simultaneous fit with the $1^+$ assumption for modes  \uppercase\expandafter{\romannumeral1} (a) and \uppercase\expandafter{\romannumeral2} (b). Dots with error bars are experimental data and the (red) solid curves show the fit model. The (blue) dashed curves are the signal component. The (violet) dotted curves show the background from the $J/\psi\rightarrow\phi\eta\eta'$ PHSP process. The (orange) dot-dashed curves represent the background from the $J/\psi\rightarrow\phi f_0(1500), f_0(1500)\rightarrow\eta\eta'$ process. The (green) long-dashed curves show the non-$\phi$ and/or non-$\etap$ backgrounds.}
\end{figure*}

Assuming the $J^P$ value of the structure to be $1^+$, the simultaneous fit
with the S-wave PHSP factor $pq$ in the partial width is performed with results 
shown in Fig.~\ref{fit_simultaneous_ss}. The mass and width of the
structure are determined to be $(2062.8\pm13.1)$ \MeV\ and
$(177\pm36)$ MeV, respectively. The
  log-likelihood value is 15595.9, with a goodness-of-fit value of
  $\chi^2/\rm{d.o.f.}$ of $16.68/26=0.64$ for mode \I\ and
  $24.36/26=0.94$ for mode \II.
The significance of the structure after considering the systematic
uncertainties (Table~\ref{t_sys_ss}) is evaluated to be 3.8$\sigma$.

\subsection{Angular distribution}

The $J^P$ assignment for
  the structure is investigated by examining the distribution of
  $|\rm cos\theta|$, where $\theta$ is the $\eta$ polar angle in the
  $J/\psi$ rest frame. If $J^P=1^-$, the decay $J/\psi\ar\eta\X$
  takes place through a P wave, neglecting the higher orbital angular
  momenta due to the closeness of the threshold, and the $|\rm cos\theta|$ is expected to follow a $1+\rm
  cos^2\theta$ distribution. If $J^P=1^+$, the above decay takes place through an S wave, where the $|\rm cos\theta|$ distribution is expected to be flat.

\begin{figure}[htbp]
\centering
{\includegraphics[width=0.4\textwidth]{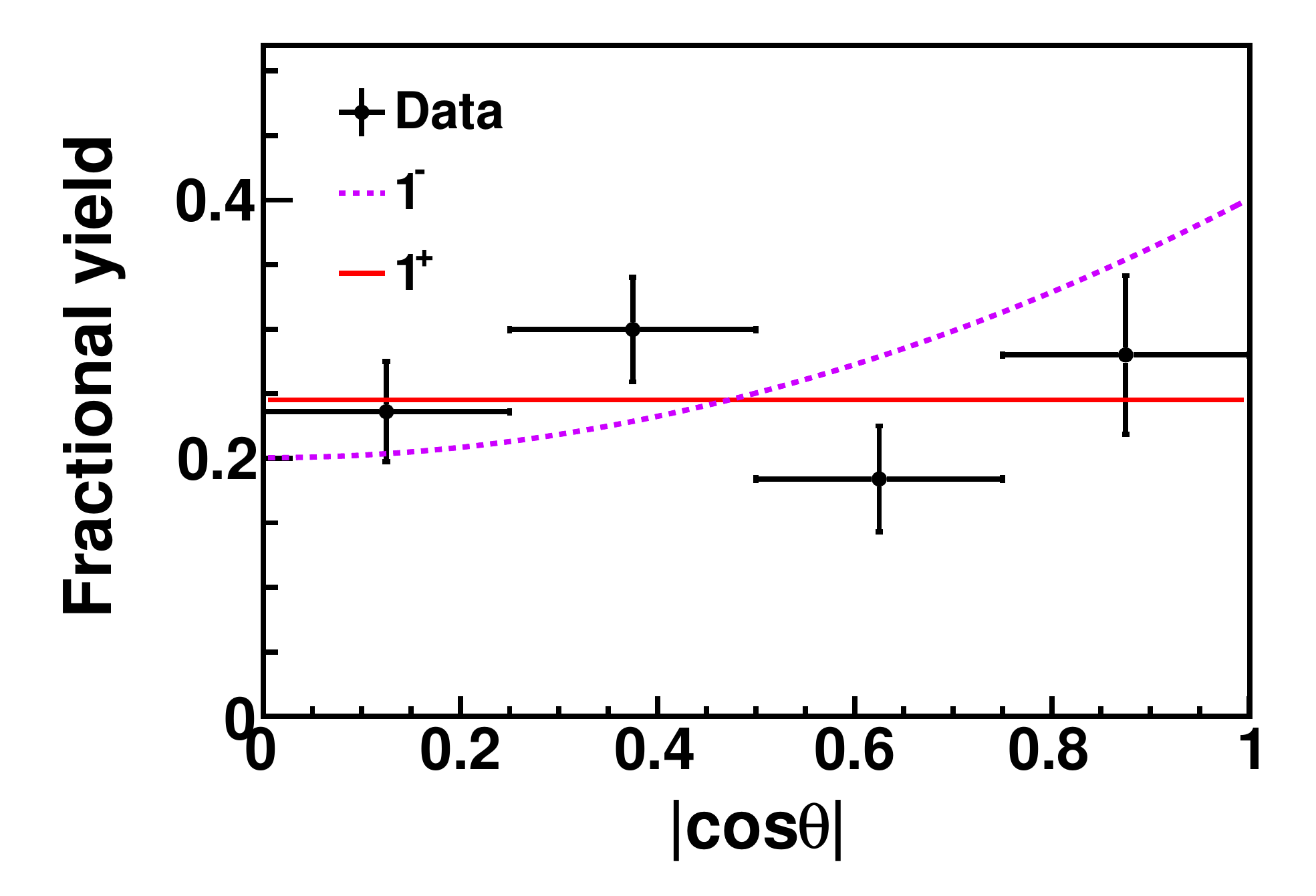}}
\caption{\label{angular_ss} Distribution of the $\eta$ polar angle in the $J/\psi$ rest frame. Dots with error bars are experimental data. The (violet) dashed curve is the fitting result with the $1^-$ assumption, and the (red) solid curve is that with the $1^+$ assumption.}
\end{figure}

The events are divided into four intervals of
  $|\rm cos\theta|$, and the total signal yield in each interval is
  obtained with the same simultaneous fit method with a $1^{+}$ assumption, as described
  above. After efficiency correction and normalization, the $|\rm cos\theta|$
  distribution of data is shown in Fig.~\ref{angular_ss}, together with the fitting results with the $1^-$ and $1^+$ assumptions. The $1^-$ assumption has
  $\chi^2/$d.o.f. value being $10.55/3=3.52$ while for the $1^+$ assumption
  it is $4.41/3=1.47$. Although the
  $\chi^2/$d.o.f. value favors the $1^+$ assumption, these two assumptions cannot clearly be distinguished due to the limited
  statistics. The $0^+$ assumption is ruled out because it violates $J^P$ conservation, and the $0^-$ assumption is rejected at 99.5$\%$ confidence level from the Pearson $\chi^2$ test. The results of simultaneous fit with $1^-$ assumption are consistent with those from $1^+$.

\subsection{Measurement of the product branching fraction}


The product branching fraction to the $\eta\phi\etap$ final state via $\X$ is 
\begin{equation}
\begin{split}
\BR(\jpsi\ar\eta\X)\times{\BR(\X\ar\phi\etap)}\\
=\frac{N_{\rm sig}}{N_{\jpsi}\BR(\eta\ar 2\g)\BR(\phi\ar{}K^+K^-)\bar{\epsilon}},
\end{split}
\end{equation}
where $N_{\rm sig}$ is the total signal yield from the two modes in the simultaneous fit; $\bar{\epsilon}$ is $\BR(\etap\ar\g\pi^+\pi^-)\epsilon_{\rm \I}+\BR(\etap\ar\eta\pi^+\pi^-)\BR(\eta\ar{}2\g)\epsilon_{\rm \II}$, where $\epsilon_{\rm \I}$ and $\epsilon_{\rm \II}$ are the detection efficiencies determined from signal MC simulation after considering the $J^P$ value of the structure and the angular distributions of the $\eta$, $\phi$ and $\etap$; the other variables have been defined before. The measured $N_{\rm sig}$ and $\BR(\jpsi\ar\eta\X)\times{\BR(\X\ar\phi\etap)}$ values for the $1^-$ and $1^+$ assumptions are summarized in Table~\ref{t_numbers}, where the uncertainties are statistical only.

    \begin{table} [tbh]
    \centering
    \caption{Measured $N_{\rm sig}$ and $\BR(\jpsi\ar\eta\X)\times{}\BR(\X\ar\phi\etap)$ values for the $1^-$ and $1^+$ assumptions.}
    \label{t_numbers}
    \begin{tabular} {lccc}
    \hline
    \hline
    $J^P$    $\quad$$\,$        & $N_{\rm sig}$ & $\BR(\jpsi\ar\eta\X)\times{}\BR(\X\ar\phi\etap)$ \\
    \hline
    $1^-$                               & $658\pm77$  & $(9.8\pm1.2)\times{10^{-5}}$ \\
    $1^+$                              & $642\pm88$  & $(9.6\pm1.4)\times{10^{-5}}$ \\
    \hline
    \hline
    \end{tabular}
    \end{table}
    

\subsection{Systematic uncertainties}

Tables \ref{t_sys} and \ref{t_sys_ss} summarise the systematic
uncertainties in the measurements of mass and width of the structure, as well as
$\BR(\jpsi\ar\eta\X)\times{}\BR(\X\ar\phi\etap)$ for the $1^-$ and $1^+$
assumptions, respectively. In case there are differences between the
uncertainties from the two modes, the more conservative values are used.

    \begin{table} [tbh]
    \centering
    \caption{Systematic uncertainties in the mass and width of the structure, as well as $\BR(\jpsi\ar\eta\X)\times{}\BR(\X\ar\phi\etap)$ (denoted as $\BR_X$ in this table) for the $1^-$ assumption.}
    \label{t_sys}
    \begin{tabular} {lccccc}
    \hline
    \hline
                & Mass & Width & \\
    Sources & (\MeV) & ($\rm MeV$) & $\mathcal{B}_X$ ($\%$)\\
    \hline
    Signal parametrization\qquad\qquad\qquad                        & 9.1 & 2 & 2.9  \\
    $f_0(1500)$                                        & 9.5 & 5 & 11.6 \\     
    PHSP assumption                              & 15.2 & 5 & 9.8 \\
    Fitting range                                       & 6.3 & 3   & 3.1   \\
    $M(\pi^+\pi^-)$ requirement               & 1.8 & 2   & 0   \\
    Extra structures                                  & 2.5 & 0   & 1.1   \\
    Momentum calibration                       & 0.7  & -    & -      \\
    Sideband regions                               & 0.9 & 2   & 0.4   \\
    MDC tracking                                     & -     & -    & 4.0   \\
    PID                                                     & -     & -    & 4.0   \\
    Photon detection                                & -     & -    & 2.4   \\
    Kinematic fit                                       & -     & -    & 3.0   \\
    Mass window for $\eta$                     & -     & -    & 0.7   \\
    Mass window for $\phi$                     & -     & -    & 1.0   \\
    Mass window for $\eta'$                    & -     & -    & 0.7   \\
    MC statistics                                      & -     & -    & 0.9   \\ 
    Branching fractions                            & -     & -    & 2.1   \\
    Number of $J/\psi$                            & -     & -    & 0.6    \\
    Total                                                   & 21.4 & 9 & 17.5  \\ 
    \hline
    \hline
    \end{tabular}
    \end{table}

    \begin{table} [tbh]
    \centering
    \caption{Systematic uncertainties in the mass and width of the structure, as well as $\BR(\jpsi\ar\eta\X)\times{}\BR(\X\ar\phi\etap)$ (denoted as $\BR_X$ in this table) for the $1^+$ assumption.}
    \label{t_sys_ss}
    \begin{tabular} {lccccc}
    \hline
    \hline
                & Mass & Width & \\
    Sources & (\MeV) & ($\rm MeV$) & $\mathcal{B}_X$ ($\%$)\\
    \hline
    Signal parametrization\qquad\qquad\qquad                        & 2.4 & 0 & 0.4  \\    
    $f_0(1500)$                                        & 2.6 & 19 & 13.4 \\ 
    PHSP assumption                              & 5.9 & 28 & 12.4 \\
    Fitting range                                       & 1.1 & 6   & 3.2   \\
    $M(\pi^+\pi^-)$ requirement               & 1.3 & 1   & 0   \\
    Extra structures                                  & 0.7 & 1   & 1.9   \\
    Momentum calibration                       & 0.7  & -    & -      \\
    Sideband regions                               & 0.7 & 1   & 0.4   \\
    MDC tracking                                     & -     & -    & 4.0   \\
    PID                                                     & -     & -    & 4.0   \\
    Photon detection                                & -     & -    & 2.4   \\
    Kinematic fit                                       & -     & -    & 2.3   \\
    Mass window for $\eta$                     & -     & -    & 0.7   \\
    Mass window for $\phi$                     & -     & -    & 1.0   \\
    Mass window for $\eta'$                    & -     & -    & 0.7   \\
    MC statistics                                      & -     & -    & 0.9   \\ 
    Branching fractions                            & -     & -    & 2.1   \\
    Number of $J/\psi$                            & -     & -    & 0.6    \\
    Total                                                   & 7.2 & 35 & 20.0  \\ 
    \hline
    \hline
    \end{tabular}
    \end{table}

The signal parametrization is changed from a
  constant-width BW function to a BW with mass-dependent
  width. The impact on the signal yield is taken as the uncertainty of
  $\BR(\jpsi\ar\eta\X)\times{}\BR(\X\ar\phi\etap)$. The pole mass
  ($m_{\rm pole}$) and pole width ($\Gamma_{\rm pole}$) are obtained
  by solving for the complex equation $P=m_{\rm pole}-i\Gamma_{\rm
    pole}/2$ for which the BW denominator is zero, and the differences between the mass and width from the nominal fit and $m_{\rm pole}$ and $\Gamma_{\rm pole}$ are considered as the uncertainties of mass and width, respectively. To obtain the uncertainties associated with the $f_0(1500)$ component of the data, the background levels in the simultaneous fit are varied by $\pm1\sigma$~\cite{c_introduction_pdg, c_introduction_exp_phi_pipi_kk}, where $\sigma$ denotes the uncertainty on the determined number of the $f_0(1500)$, and the maximum changes in the fit results are regarded as uncertainties. We also vary the non-resonant $\phi\eta\etap$ background levels by $\pm1\sigma$, and take the largest influences on the fit results as the uncertainties due to the PHSP assumption. We vary the range of the simultaneous fit by 5$\%$ and take the largest deviations of the fitting results as uncertainties. To obtain the uncertainties due to the $M(\pip\pim)$ requirement for mode \uppercase\expandafter{\romannumeral1}, it is relaxed from 0.87 to 0.90 \GeV\ and the effects on the fitting results are considered as uncertainties. The two possible extra structures around 2.3 \GeV\ in Figs.~\ref{fit_simultaneous} (b) and \ref{fit_simultaneous_ss} (b) are considered. Following the procedure in Ref.~\cite{c_introduction_besiii}, we use BW functions convolved with a resolution function to describe them and the corresponding significances are determined to be less than 1.1$\sigma$, and they are not considered in the nominal result. However, their impacts on the fitting results are taken as systematic uncertainties. The difference between the fitted $\eta$ mass and that from the PDG~\cite{c_introduction_pdg} is taken as the uncertainty due to momentum calibration. The descriptions of other items are included in Table \ref{t_sys_2}. The total systematic uncertainties are the quadrature sums of the individual contributions, assuming they are independent.


\section{Summary and discussion}

In summary, using $(1310.6\pm7.0)\times{10^6}$ $J/\psi$ events
collected with the BESIII detector, we report the observation and study of the  
process $J/\psi\rightarrow\phi\eta\eta'$. Its branching fraction, including all possible
intermediate states, is determined to be
$(2.32 \pm 0.06 \pm 0.16) \times 10^{-4}$. Evidence for a structure denoted as $\X$
in the $\phi\eta'$ mass spectra in two dominant $\eta'$
decay modes is reported, and a simultaneous fit is
performed. Assuming the $J^P$ value of the structure to be $1^-$, the
significance of the structure is evaluated to be 4.4$\sigma$; the mass
and width are determined to be
$(2002.1 \pm 27.5 \pm 21.4)$ \MeV\ and $(129 \pm 17 \pm 9)$ MeV,
respectively; the product branching fraction
$\BR(\jpsi\ar\eta\X)\times{}\BR(\X\ar\phi\etap)$ is measured to be
$(9.8 \pm 1.2 \pm 1.7)\times 10^{-5}$. The mass of the structure is over 5$\sigma$ away from that of the $\Y$ in the PDG~\cite{c_introduction_pdg}, suggesting the structure might not be the $\Y$. For a $1^+$ assumption, the significance is evaluated to be
3.8$\sigma$; the mass and width are determined to be
$(2062.8 \pm 13.1 \pm 7.2)$ \MeV\ and $(177 \pm 36 \pm 35)$ MeV,
respectively; the product branching fraction
$\BR(\jpsi\ar\eta\X)\times{}\BR(\X\ar\phi\etap)$ is measured to be
$(9.6 \pm 1.4 \pm 2.0)\times 10^{-5}$. The
  angular distribution is studied and the $1^-$ and $1^+$ assumptions
  cannot clearly be distinguished due to the limited statistics. No
meson candidate in the PDG has mass, width and $J^P$ values that are compatible with
the structure. More studies with a larger $\jpsi$ data sample in the future might help to better understand the structure, including a $J^P$ determination and precise measurements of the mass, width, and product branching fraction. 


\acknowledgments

This work is supported in part by National Key Basic Research Program of China under Contract No. 2015CB856700; National Natural Science Foundation of China (NSFC) under Contracts Nos. 11335008, 11425524, 11625523, 11635010, 11735014; the Chinese Academy of Sciences (CAS) Large-Scale Scientific Facility Program; the CAS Center for Excellence in Particle Physics (CCEPP); Joint Large-Scale Scientific Facility Funds of the NSFC and CAS under Contracts Nos. U1532257, U1532258, U1832207; CAS Key Research Program of Frontier Sciences under Contracts Nos. QYZDJ-SSW-SLH003, QYZDJ-SSW-SLH040; 100 Talents Program of CAS; the Institute of Nuclear and Particle Physics and Shanghai Key Laboratory for Particle Physics and Cosmology; German Research Foundation DFG under Contracts Nos. Collaborative Research Center CRC 1044, FOR 2359; Istituto Nazionale di Fisica Nucleare, Italy; Koninklijke Nederlandse Akademie van Wetenschappen (KNAW) under Contract No. 530-4CDP03; Ministry of Development of Turkey under Contract No. DPT2006K-120470; National Science and Technology fund; The Swedish Research Council; U. S. Department of Energy under Contracts Nos. DE-FG02-05ER41374, DE-SC-0010118, DE-SC-0010504, DE-SC-0012069; University of Groningen (RuG) and the Helmholtzzentrum fuer Schwerionenforschung GmbH (GSI) Darmstadt; Institute for Basic Science (Korea) under project code IBS-R016-D1.


\end{document}